\documentclass[onecollarge]{svjour2}      
\smartqed  
\usepackage{amsmath}
\usepackage{amssymb}
\usepackage{graphicx}
\usepackage{txfonts}      

\journalname{Acta Mech.}

\begin{document}

\title{Comment on ``On a class of exact solutions of the equations of motion of a second grade fluid'' by C.\ Fetec\u{a}u and J.\ Zierep (Acta Mech.\ 150, 135--138, 2001)}


\author{Ivan C.\ Christov \and C.\ I.\ Christov}


\institute{
Ivan C.\ Christov (corresponding author) \at
Department of Engineering Sciences and Applied Mathematics, Northwestern University, Evanston, IL 60208-3125, USA\\
Tel.: +1-847-491-3555\\
Fax: +1-847-491-2178\\
\email{christov@u.northwestern.edu}
\and
C.\ I.\ Christov \at
Department of Mathematics, University of Louisiana at Lafayette, Lafayette, LA 70504-1010, USA\\
\email{christov@louisiana.edu}
}

\date{Received: date / Accepted: date}

\maketitle

\paragraph{Introduction}
In a 2001 article, Fetec\u{a}u and Zierep~\cite{FZ01} considered Stokes' first problem for a second grade (SG) fluid, unaware that it had already been solved (correctly) by Puri~\cite{P84} in 1984. These authors used the Fourier sine transform to obtain, what they believed to be, the exact solution of the resulting initial-boundary-value problem (IBVP). Unfortunately, due to their incorrect computation of the distributional derivative of the Heaviside function, an omission of a critical term in the subsidiary equation occurred, an elementary mistake (gracefully explained in \cite{J10}) apparently not uncommon in the literature. Therefore, the solution given in \cite{FZ01} is, generally speaking, incorrect.

The aim of the present comment is to point out and correct this error. Hence, employing the same notation convention used in \cite{FZ01}, we begin by restating the IBVP considered therein:
\begin{subequations}\begin{align}
&(\nu + \alpha\partial_t)\partial_y^2	v(y,t) = \partial_t v(y,t),\qquad (y,t)\in(0,\infty)\times(0,\infty);\label{eq:pde}\\
&v(y,0) = 0,\quad y>0;\qquad v(0,t) = V(t),\quad v(y,t),\; \partial_y v(y,t) \to 0 \;\;\text{as}\;\; y\to\infty,\qquad t>0.\label{eq:bc}
\end{align}\label{eq:ibvp}\end{subequations}
Here, $\alpha$ and $\nu$ are positive constants. Note that $V(t)$ as written above and in \cite{FZ01} is an ambiguous notation because the function is \emph{not} arbitrary. For this IBVP to be Stokes' first problem, the plate must move \emph{suddenly} at $t=0^+$. Therefore, $V(t)$ \emph{must} be such that $V(t) = \tilde V (t) \theta(t)$, where $\theta(\cdot)$ denotes the Heaviside unit step function and $\tilde V(t)$ is some smooth function. (Stokes himself made this distinction clear \cite[pp.~101]{S19}.) Consequently, the correct expression in lieu of $V'(t)$ in \cite[Eq.~(2.3)]{FZ01} is $\tilde V'(t)\theta(t) + \tilde V(t)\delta(t)$, where $\delta(\cdot)$ denotes the Dirac delta function. In the case of $\tilde V(t) \sim t^n$ as $t\to0$, for some integer $n>1$, the singularity is ameliorated and by pure luck the proper answer may be obtained despite the erroneous derivation.

\paragraph{Corrected solution}
\label{sec:2}
The simplest case is that of \emph{impulsive} plate motion with constant velocity: $v(0,t)=V_0\theta(t)$. Thus, $V'(t)$ in \cite{FZ01} becomes $V_0\delta(t)$ and it  follows that the solution of the subsidiary equation (i.e., \cite[Eq.~(2.3)]{FZ01}) is, in this case, given by 
\begin{equation}
\bar{v}_{s}(\xi,s) = 2V_0\left\{\frac{\nu \xi}{s[s(1+\alpha\xi^2)+\nu \xi^2]}+ \frac{\alpha\xi}{s(1+\alpha\xi^2)+\nu \xi^2}\right\},
\label{eq:transform_soln}
\end{equation}
which we obtained using the Laplace transform.  Here, a bar over a quantity denotes its image  in the Laplace transform domain and $s$ denotes the Laplace transform parameter. Note that we employ the Fourier sine transform convention from \cite{CB78}.

Using partial fractions, a standard table of Laplace  inverses, and the definition of the inverse sine transform,  Eq.~\eqref{eq:transform_soln} is easily inverted to yield the time-domain solution
\begin{equation}
v(y,t) = V_0\theta(t)\left[1-\frac{2}{\pi}\int_{0}^{\infty}\frac{\sin(\xi y)}{\xi} \exp\left(\frac{-\nu \xi^2t}{1+\alpha\xi^2}\right)\,\mathrm{d}\xi+ \frac{2\alpha}{\pi}\int_{0}^{\infty}\frac{\xi \sin(\xi y)}{1+\alpha\xi^2}\exp\left(\frac{-\nu \xi^2 t}{1+\alpha\xi^2}\right)\,\mathrm{d}\xi\right].
\label{eq:soln}
\end{equation}
Here, we observe that the second integral in Eq.~\eqref{eq:soln}, which represents the contribution from the second quotient in Eq.~\eqref{eq:transform_soln},  is \emph{not} present in \cite[Eq.~(2.6)]{FZ01}; consequently, the latter is incorrect.  This error appears to be due to the authors of \cite{FZ01} taking  $V'(t)$ as zero, rather than its correct value $V_0\delta(t)$. Equation~\eqref{eq:soln} above is, therefore, the \emph{correct} exact solution, based on the Fourier sine and Laplace transforms, to Stokes' first problem for a SG fluid.

The same mistake is committed when going from Eq.~(3.3) to Eq.~(3.4) in \cite{FZ01}, however, we leave it as an exercise to the reader to find the correct solution in this case.

An additional error transpires in the integration of the subsidiary equation \cite[Eq.~(2.3)]{FZ01}. Though it is claimed that \cite[Eq.~(2.4)]{FZ01} is its solution, it is clear that, by the integrating factor method, which appears to be what was attempted, such a solution cannot be obtained. The same mistake occurs in the derivation of \cite[Eq.~(3.3)]{FZ01}. We find it unnecessary to supply corrected expressions for these intermediate formul\ae\ because they already suffer the much graver error, committed in the taking the derivative of the Heaviside function, that we corrected above.

\paragraph{A simple finite-difference scheme}
To provide an independent check on the corrected transform solution given in Eq.~\eqref{eq:soln}, we also solve the IBVP \eqref{eq:ibvp} numerically. We define the (uniform) spatial- and temporal-step sizes $\Delta y := L/(M-1)$ and $\Delta t := t_f/(K-1)$, where $M\ge2$ and $K\ge 2$ are integers and now $(y,t)\in(0,L)\times(0,t_f]$. Then, letting $\mathfrak{v}_j^n \approx v(y_j,t^n)$ be the approximation to the exact solution on the grid, where $y_j := j\Delta y$ ($0\le j \le M-1$) and $t^n := n\Delta t$ ($0\le n \le K-1$), we may discretize Eq.~\eqref{eq:pde} as follows:
\begin{equation}
\delta_{t+}\mathfrak{v}^{n}_j - \alpha \delta_{t+}\delta_{y+}\delta_{y-}\mathfrak{v}^{n}_j = \nu  \frac{1}{2}\left(\delta_{y+}\delta_{y-}\mathfrak{v}^{n+1}_j + \delta_{y+}\delta_{y-}\mathfrak{v}^{n}_j\right).
\label{eq:scheme}
\end{equation}
Here, $\delta_{t+}$ is the forward temporal difference operator and $\delta_{y+}$ and $\delta_{y-}$ are, respectively, the forward and backward spatial difference operators \cite[\S3.3]{S04}. It is a straightforward, though lengthy, calculation (see, e.g., \cite{S04}) to show that this implicit, two-level Crank--Nicolson-type discretization is unconditionally stable and has truncation error $\mathcal{O}[(\Delta t)^2 + (\Delta y)^2]$. The boundary conditions are implemented as
\begin{equation}
\mathfrak{v}_0^n = \begin{cases}0, &n=0;\\ V_0, &1\le n\le K-1. \end{cases};\qquad \mathfrak{v}_{M-1}^n = 0,\quad 0\le n\le K-1.
\end{equation}
For appropriately chosen $L\gg1$, the front does not reach the $y=L$ boundary for any $t\in(0,t_f]$, and so this is the ``numerical infinity.''

For the computations shown below, we took $M=5000$, $K=1000$ and $L = 20$ to obtain a highly-accurate solution that we can compare to Eq.~\eqref{eq:soln}. {\sc Matlab}'s built-in Gaussian elimination algorithm was used to invert the symmetric tridiagonal matrix resulting from this discretization.

\paragraph{Illustrated example}
Here we compare the correct solution given in Eq.~\eqref{eq:soln} and the numerical solution by the difference scheme in Eq.~\eqref{eq:scheme} to the incorrect solution given in \cite[Eq.~(2.6)]{FZ01}. Clearly, Eq.~\eqref{eq:soln} and the numerical solution are in excellent agreement. Meanwhile, the solution from \cite[Eq.~(2.6)]{FZ01} does not even satisfy the boundary condition $\lim_{y\to0^+} v(y,t) = V_0$ $(t >0)$.
\begin{figure}[!ht]
\centerline{\includegraphics[width=0.45\textwidth]{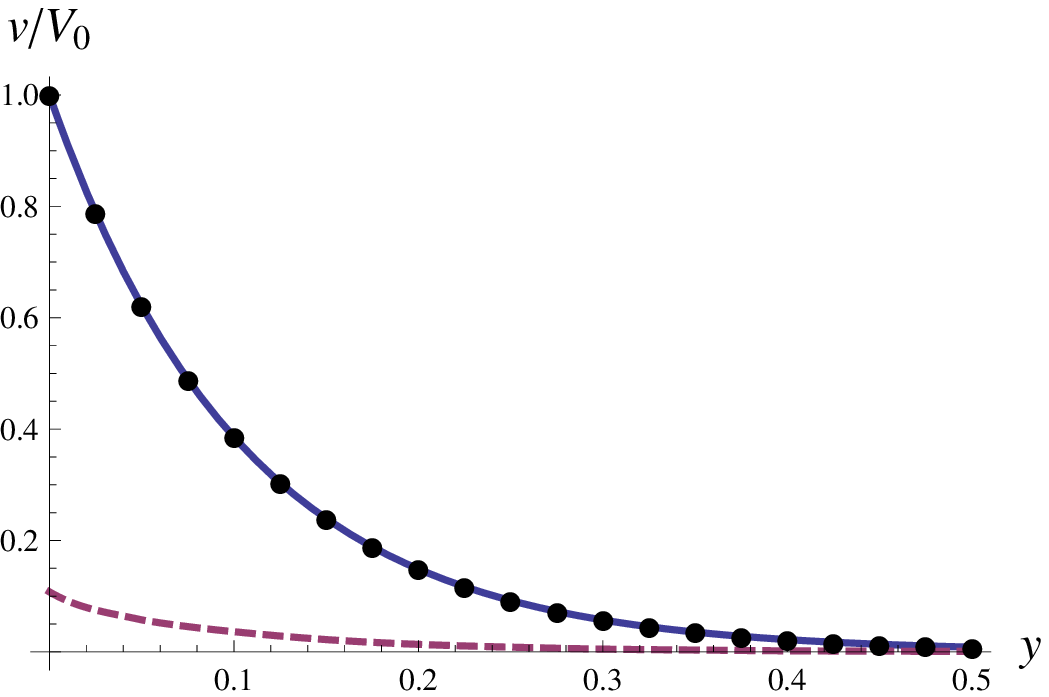}\hspace{1cm}\includegraphics[width=0.45\textwidth]{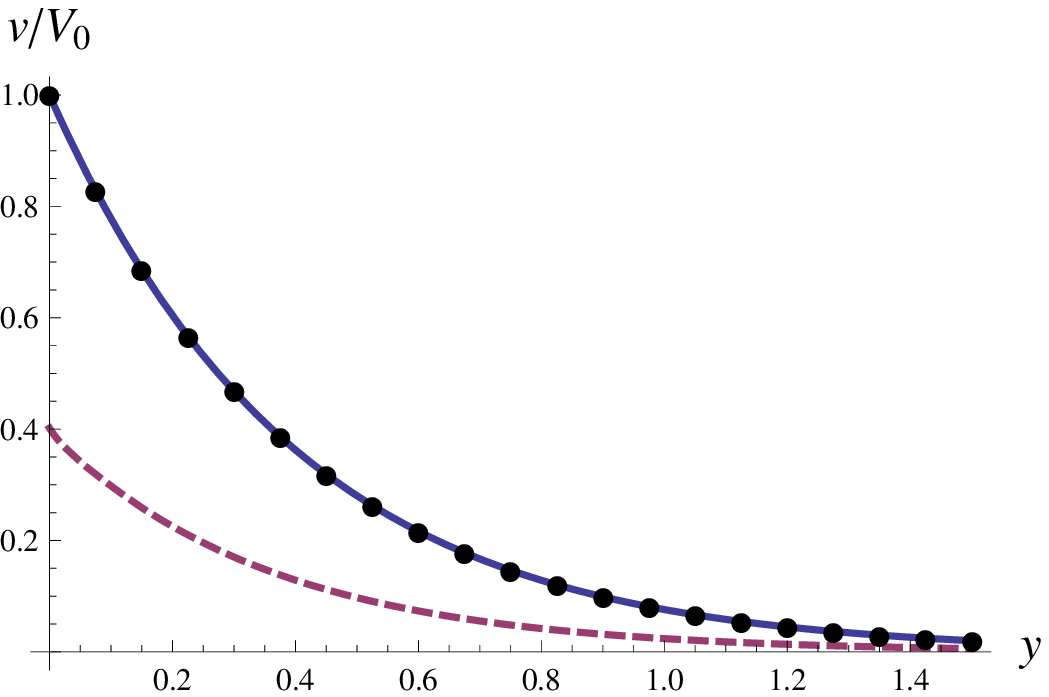}}
\caption{$v$ vs.\ $y$ for (left) $\nu = 0.1$, $\alpha = 0.01$, $t=0.01$ and (right) $\nu = 0.1$, $\alpha = 0.1$, $t=0.5$. Dashed: incorrect solution \cite[Eq.~(2.6)]{FZ01}; solid: correct solution [Eq.~\eqref{eq:soln} above]; dots: numerical solution [using the scheme in Eq.~\eqref{eq:scheme}].}
\label{fig:soln}
\end{figure}

\paragraph{Conclusion}
It is regrettable that this error has now spread throughout the literature on non-Newtonian fluid mechanics, ``infecting'' a number of journals over the last nine years. The same error also appears in work on Stokes' first problem for relaxing fluids \cite{CJ09}. This means that one must be very careful when applying repeated integral transforms to problems with mixed derivatives and discontinuous initial and/or boundary conditions.

It is curious to note that if the contrived source term $2V_0\alpha\delta(t)\delta'(y)$ were added to the left-hand side of Eq.~\eqref{eq:pde}, the error in computing $V'(t)$ would be ``cancelled'' and \cite[Eq.~(2.6)]{FZ01} would be the solution to this new problem. Similarly, if the boundary condition in Eq.~\eqref{eq:bc} were changed to $v(0,t) = (1 - \mathrm{e}^{-\nu t /\alpha})V_0\theta(t)$, then \cite[Eq.~(2.6)]{FZ01} coincides with the well-know solution to this problem from the theory of unsteady flow in fissured rocks \cite[Eq.~(5.9)]{BZK60}. This is the kind of lucky occurrence we alluded to in the Introduction. It is noteworthy that \cite{BZK60} is a seminal paper with several hundred citations, yet it somehow eluded the authors of \cite{FZ01}.

In addition, it is of interest to note that if we set 
\begin{equation}
u=v/V_0,\qquad Y=y,\qquad T =\nu t, \qquad l^2=\alpha,\qquad \eta=\frac{\xi^2}{1+l^2\xi^2},
\end{equation}
then Eq.~\eqref{eq:soln}  can be re-expressed as
\begin{equation}
u(Y,T)=\theta(T)
\left[1-\frac{1}{\pi}\int_{0}^{l^{-2}}\!{\rm e}^{-\eta T} \sin\left( \frac{Y}{l}\sqrt{\frac{\eta}{l^{-2}-\eta}}\,\right)\!\frac{{\rm d}\eta}{\eta}\right].
\end{equation}
The latter is equivalent to the $b=0$ special case of \cite[Eq.~(10)]{J10}, wherein the \emph{correct} porous medium version of IBVP \eqref{eq:ibvp} is solved, a result which can also be obtained using \emph{only} the Laplace transform with respect to time \cite{JP03,P84}. This means that it is possible to prevent the error corrected herein by avoiding the use of multiple integral transforms simultaneously.

Finally, the assumption that $\alpha$ is positive in Eq.~\eqref{eq:pde} is an important one because $\alpha < 0$ renders the rest state of the SG fluid unstable \cite{J81} and the problem ill-posed. Yet, certain theoretical arguments and experimental measurements involving polymer solutions suggest that $\alpha$ may be negative \cite{T00}. All this serves to point out the approximate nature of the SG fluid model \cite{A02}. In fact, the sign of $\alpha$ was not addressed in \cite{FZ01}, and the correction of the mathematical solution we have presented should not be construed as resolving any fundamental difficulties with the mechanics of SG fluids.

\begin{acknowledgements}
I.C.C.\ was supported by a Walter P. Murphy Fellowship from the Robert R.\ McCormick School of Engineering and Applied Science at Northwestern University.
\end{acknowledgements}

\columncase{}{\newpage}

\end{document}